\documentclass[12pt,a4paper]{article}
\usepackage[T1]{fontenc}
\usepackage[english]{babel}
\usepackage{graphics}
\usepackage{setspace}

\doublespace

\begin{document}

\title{Monte Carlo simulation of metal deposition on foreign substrates.}
\author{M.C. Gim\'{e}nez \thanks{
Corrresponding author: M.C. Gim\'{e}nez, e-mail: cecigime@unsl.edu.ar}, A. J. Ramirez-Pastor \\
%EndAName
Laboratorio de Ciencias de Superficies y Medios Porosos,\\
Departamento de F\'{\i}sica,\\
Facultad de Ciencias F\'{\i}sico-Matem\'aticas y Naturales,\\
Universidad Nacional de San Luis, CONICET,\\
Chacabuco 917, 5700, San Luis, Argentina\\
\\
\and E. P. M. Leiva \\
%EndAName
Unidad de Matem\'atica y F\'{\i}sica, Facultad de Ciencias Qu\'{\i}micas,\\
Universidad Nacional de C\'{o}rdoba, 5000\\
C\'{o}rdoba, Argentina\\
\\
}
\maketitle

\begin{abstract}
\bigskip

The deposition of a metal on a foreign substrate is studied by means of
grand canonical Monte Carlo simulations and a lattice-gas model with pair
potential interactions between nearest neighbors. The influence of
temperature and surface defects on adsorption isotherms and differential
heat of adsorption is considered. The general trends can be explained in
terms of the relative interactions between adsorbate atoms and substrate
atoms. The systems $Ag/Au(100)$, $Ag/Pt(100)$, $Au/Ag(100)$ and $Pt/Ag(100)$
are analyzed as examples.

\bigskip

\end{abstract}

\section{Introduction}

The study of the adsorption of particles on surfaces is important from the
point of view of surface science and due to its potential applications in
nanotechnology and catalysis.

From the electrochemical point of view, it is of great interest to study the
electrodeposition of a metal onto a single crystalline surface of a foreign
metal \cite{Kolb, Trasatti, Lorenz_Libro, Lorenz_2000}. When this occurs at
potentials more positive than those predicted from the Nernst equation, the
process is denominated underpotential deposition (UPD) \cite{Kolb,
Lorenz_Libro, Leiva_UPD_1, Leiva_UPD_2, Lorenz_2004}. This can be
intuitively understood considering that, in general, UPD is observed when
for the adsorbate it is more favorable to be deposited on the considered
substrate than on a surface of its same nature. When the opposite occurs,
the observation of another phenomenon called overpotential deposition (OPD)
is expected , which implies that the adsorption of the adsorbate on the
foreign substrate is less favorable than that predicted by the Nernst
equation.

It is clear that a complete analysis of the UPD phenomenon is a quite
difficult subject because of the complexity of the involved systems. For
this reason, the understanding of simple models with increasing complexity
might be a help and a guide to establish a general framework for the study
of this kind of systems. Pioneering work using Monte Carlo simulations to
study underpotential multisite adsorption has been undertaken by Van Der
Eerden et al. \cite{Van der Eerden}.In this context, the present work tries
to contribute to the comprehension of some essential characteristics of the
metal deposition by means of a very simple model. For this purpose, we
present simulations of metal adsorption on metal surfaces, employing a
lattice-gas model with pair potential interactions characterized by a few
parameters. Depending on the particular values assigned to the parameters,
we can represent different metals involved. As illustrative examples we have
simulated the following systems: $Ag/Au(100)$ (that means: adsorption of
silver atoms on gold (100) surfaces), $Ag/Pt(100)$, $Au/Ag(100)$ and $%
Pt/Ag(100)$.

In previous works \cite{MC1, MC2} we have used a more complicated model for
the calculation of energies, employing the embedded atom method, which is
more appropriate for metals in the sense that takes into account many-body
interactions. The cost of introducing this precursor model is the lack of
some experimental features presented by real systems. However, there are
important physical motivations to pay this cost: a) this contribution allows
to identify and characterize the most prominent features of this particular
process; b) the study offers a general theoretical framework in which this
kind of process can be studied and c) the conclusions obtained are
interesting. The outline of the paper is as follows: In Section 2 we
describe the lattice-gas model along with the simulation scheme. In Section
3 we present the results. Finally, the general conclusions are given in
Section 4.

\section{Model and simulation method}

\subsection{Lattice model}

Lattice models for computer simulations are of widespread use in studies of
adsorption on surfaces, because they allow dealing with a large number of
particles at a relatively low computational cost \cite{Nicholson}.

If the crystallographic misfit between the involved atoms is not important,
it is a good approximation to assume that the adatoms adsorb on defined
discrete sites on the surface, given by the positions of the substrate
atoms. This is the case of the very well studied system $Ag/Au(100)$ \cite%
{Ag/Au_Garcia, Ag/Au_Marijo, Ag/Au_Garcia2, Ag/Au_Hara}.

Square lattices with periodical boundary conditions are used in the present
work to represent the surface of the electrode. Each lattice point
represents an adsorption site for an adsorbate or a substrate atom. The
former may adsorb or desorb on each randomly selected site, while the latter
may only move on the surface jumping from the selected site to a neighboring
one. In this way, our model corresponds to an open system for one of its
components, that is, the adsorbate. This has physical correspondence with
the setup of the electrochemical experiment, where only the metal in
equilibrium with its cations in solution may dissolve or be deposited in the
potential range considered.

Concerning the adatoms of the same kind of the substrate, some
considerations must be made regarding the existence of surface defects.
These atoms may in principle move freely on the surface and minimize the
free energy of the system by a number of mechanisms. For example, isolated
substrate atoms may heal defects through their incorporation to a defective
cluster, or small substrate islands may dissolve to join large ones, like
shown in previous simulation work by Stimming and Schmickler \cite{S&S}.
Thus, different substrate structures can be imposed as initial conditions
for each simulation. In the present case, we employ islands of substrate
atoms on the surface obtained by means of simulated annealing techniques 
\cite{MC1}. This was undertaken in order to emulate some of the defects that
can be found on a real single crystal surfaces, like kink sites, steps, etc.

On purely energetic grounds, metal electrodeposition on clean metal surfaces
with islands should take place according to the following sequence:

i)Decoration of the lower part of island edges. This is so because the
binding energy is the lower(more negative), the higher the coordination of
the site where a metal atom becomes adsorbed.

ii)Deposition on the rest of the flat surface around the islands.

iii)Some nucleation should occur at the top of the islands. This requires a
more negative potential than that for island decoration(i) or flat surface
filling(ii) because the border of the growing nuclei makes them less stable
than a growing monolayer (nucleation overpotential)

iv)The border of the islands at the top should be finally decorated. These
sites are particularly unstable due to their low coordination.

In the present work we are mainly interested in island decoration and the
possible mixing between island atoms and depositing atoms following
immediately after this decoration. This involves steps i) and ii) described
above. Steps iii) and iv) could be of course simulated but this will not be
considered in the present work.

\subsection{Energy calculations}

The energy $E$ of the system is related with the classical lattice-gas model
with pair potential interactions between nearest neighbors:

{\centering}

{\ 
\begin{equation}
E=\sum_{\alpha }(\epsilon _{1}\delta _{c_{\alpha },1}+\epsilon _{2}\delta
_{c_{\alpha },2})+\sum_{\langle \alpha ,\beta \rangle }\left[ J_{11}\delta
_{c_{\alpha },1}\delta _{c_{\beta },1}+J_{22}\delta _{c_{\alpha },2}\delta
_{c_{\beta },2}+J_{12}(\delta _{c_{\alpha },1}\delta _{c_{\beta },2}+\delta
_{c_{\alpha },2}\delta _{c_{\beta },1})\right]  \label{2}
\end{equation}%
}

\noindent where the symbol $\delta $ represents the Kronecker delta and $%
c_{\alpha }$ indicates the occupation number of the $\alpha $ site
[corresponding to a pair of $(x,y)$ coordinates]. Unlike the classical
lattice-gas model, the occupation number may assume three different values: $%
0$, $1$ or $2$, if the corresponding site is empty, occupied by an atom of
the same kind of the substrate or occupied by an adsorbate atom,
respectively. The notation $\langle \alpha ,\beta \rangle$ implies a sum
taking account of all pairs of nearest neighbors. $\epsilon _{i}$ is the
adsorption energy of a particle of type $i$ and $J_{ij}$ are the pairwise
interactions between nearest neighbors.

The values of the parameters considered here for the different systems are
summarized in table $1$. These values were obtained according to
calculations employing the embedded atom model \cite{Daw-Baskes,MC1, MC2}.
In a previous work \cite{MC2}, we have calculated the adsorption energies
corresponding to the different environments of the adsorption site, taking
into account first, second and third neighbors. Some of the possible
configurations were ilustrated in Fig. $1$ and corresponding energies were
indicated in table $1$ of reference \cite{MC2}. The values employed in the
present work for $\epsilon _{i}$ were based on the adsorption energies
corresponding to configuration $1$ (that is, adsorption on a site without
neighbors). The values employed for the parameters $J_{ij}$ were based on
the others configurations ($2$, $3$, etc., which correspond to different
occupation of neighboring sites) assuming pair potential interactions
between nearest neighbors and taking an average value.

\subsection{Grand Canonical Monte Carlo}

Square lattices of $M=L\times L$ adsorption sites with periodical boundary
conditions are used here to represent the surface (in this case, $L=100$).
In the case of adsorption isotherms on clean surfaces, the initial state
consists of $2000$ adsorbate atoms distributed at random (that makes a
coverage degree of $0.2$). For the case of adsorption in the presence of
surface defects, the initial state is characterized by $N_{d}=1000$ atoms of
the same nature as that of the substrate forming islands previously
generated by simulated annealing \cite{MC1, MC2}, and $2000$ adsorbate atoms
distributed at random. The islands of substrate atoms are useful to emulate
surface defects like kink and step sites.

The simulation consists in the realization of a certain number of Monte
Carlo Steps (MCS) in order to equilibrate the system and then another set of
MCS evaluation, from time to time, the thermodynamic quantities of interest
(like energy of the system or coverage degree) in order to obtain the
average value. This is performed for fixed values of temperature and
chemical potential (Grand Canonical ensemble).

Each MCS implies the realization of $M$ trials. Each trial consists on the
random selection of one lattice site, with the realization of one of the
three following processes:

\begin{itemize}
\item[a)] If the site is empty (occupation $0$), the creation of an
adsorbate atom is attempted (yielding occupation $2$).

\item[b)] If the site is occupied by an adsorbate atom (occupation $2$), its
desorption is attempted (occupation $0$).

\item[c)] If the site is occupied by a substrate atom (occupation $1$) a
nearest neighbor site is selected at random. If this new site is empty, \ a
move to this position is tried.
\end{itemize}

In the three cases, the change is accepted with probability:

{\centering}

{\ 
\begin{equation}
P=\min \{1,\exp [-\frac{\Delta E-\mu \Delta N_{a}}{k_{B}T}]\}  \label{3}
\end{equation}%
}

\noindent where $\Delta E$ ($\Delta N_{a}$) represents the difference
between the energies (number of adsorbed particles) of the final and initial
states, $k_{B}$ is the Boltzmann constant, $T$ is the temperature and $\mu $
is the chemical potential. Note that the value of $\Delta N_{a}$ is $+1$ in
case $a)$, $-1$ in case $b)$ and $0$ in case $c)$.

The algorithm described above has some differences with the algorithm
employed in Refs. \cite{MC1, MC2}. In the present case, only the substrate
atoms can move between adjacent sites, although this possibility is
indirectly contemplated for adsorbate atoms in the case of desorption from
one site and adsorption in a neighboring site. This difference is not
essential, the main target was to save computational time.

\subsection{Some definitions}

\subsubsection{Coverage degree}

The coverage degree, $\theta (\mu )$, which is defined as the number of
sites occupied by an adsorbate atom divided by the number of available sites
(all adsorption sites that are not occupied by a substrate atom), is
obtained as a simple average:

\begin{equation}
\theta(\mu) = \frac{1}{M-N_d} \sum_{\alpha} \langle \delta_{c_{ \alpha },2}
\rangle = \frac{\langle N_a \rangle}{M-N_d}
\end{equation}

\noindent where $\langle N_{a}\rangle $ is the mean number of adsorbate
atoms on the surface and $\langle \cdots \rangle $ denotes the time average
over the Monte Carlo simulation runs.

In the case of adsorption in the presence of surface defects, that is, when
some substrate islands are present, we also define some partial coverages,
like the coverage degree of steps sites, $\theta_s$, and the coverage degree
of kink sites, $\theta_k$:

\begin{equation}
\theta_{s} = \frac{ \langle N_{s, o} \rangle}{ \langle N_{s, t} \rangle}
\end{equation}

\begin{equation}
\theta_{k} = \frac{ \langle N_{k, o} \rangle}{\langle N_{k, t} \rangle}
\end{equation}

\noindent where $\langle N_{s,t}\rangle $ ( $\langle N_{k,t}\rangle $ ) is
the average number of step (kink) sites on the surface and $\langle
N_{s,o}\rangle $ ( $\langle N_{k,o}\rangle $ ) is the average number of step
(kink) sites occupied by an adsorbate atom.

\subsubsection{Differential heat}

We calculate the differential heat of adsorption, defined as \cite{Hall}:

\begin{equation}
q_{d}=-\frac{\partial \langle E\rangle }{\partial \langle N_{a}\rangle }
\end{equation}

\noindent where $\langle E\rangle $ is the mean energy of the adlayer.

The physical interpretation of this quantity is the energy associated with
removing a particle from the surface at each moment.

In the case of computer simulations, the most appropriate way of performing
the calculation of this value is by means of averaged quantities. We employ
the following formulation \cite{Nicholson, Bulnes}:

\begin{equation}
q_{d}=-\frac{\langle EN_{a}\rangle -\langle E\rangle \langle N_{a}\rangle }{%
\langle N_{a}^{\ 2}\rangle -\langle N_{a}\rangle ^{2}}
\end{equation}

\subsubsection{Quantification of island disintegration}

As discussed below, for some systems the substrate islands are found to
disintegrate upon adatom deposition. In order to quantify this phenomenon,
we define some (normalized) quantities characterizing the disintegration of
substrate islands present on the surface when the adsorbate particles
adsorb, as follows:

\begin{itemize}
\item Fraction number of $1-0$ pairs:

\begin{equation}
NP_{1-0}=\frac{\sum_{<\alpha ,\beta >}\delta (1,c_{\beta })\times \delta
(0,c_{\alpha })}{cN_{d}}
\end{equation}

\noindent where the sum yields the number of adjacent substrate atom-empty
site pairs. $c$ is the lattice connectivity (in this case $c=4$).

\item Fraction number of $1-2$ pairs:

\begin{equation}
NP_{1-2}=\frac{\sum_{<\alpha ,\beta >}\delta (1,c_{\beta })\times \delta
(2,c_{\alpha })}{cN_{d}}
\end{equation}

\noindent where the sum yields the number of adjacent substrate-adsorbate
pairs.

\item Total number of \ $1-X$ pairs, with $X=0,2$, that we define from:

\begin{equation}
NP_{T}=NP_{1-0}+NP_{1-2}  \label{NPT}
\end{equation}

\noindent which corresponds to the fraction of sites surrounding substrate
atoms that are not occupied by particles of the same kind (type 1). The
resulting quantity, further on called "fraction of island disintegration",
is given by the number of substrate atoms/adsorbate atoms and substrate
atoms/empty sites pairs.
\end{itemize}

The maximum value of $NP_{T}$ will be reached in the case where the islands
are completely disintegrated, that is, when there is a minimum of
neighboring substrate-substrate pairs. In principle, the upper value of $%
NP_{T}$ could be $1$, in the case where no substrate-substrate pairs occur
at the maximum coverage by the adsorbate. However, if we only take into
account statistical considerations, it is to be expected that the
probability of finding a neighbor of the same species ($1$) should be given
by the fraction $N_{d}/M$, which in the present simulation is fixed to the
value equal to $0.1$. In other words, it is expected that in the limit of
high temperatures where entropic effects prevail, and for large $\theta $,
we should get that $NP_{T}\longrightarrow 0.9.$

For a given distribution of well equilibrated substrate islands, the minimum
value of $NP_{T}$ will be given by the fraction of pairs involving atoms
that belong to the edge of the island at beginning of the simulation. If the
islands do not disintegrate at all, remaining intact, $NP_{T}$ should remain
the same all over the simulation. On the other hand, if the islands
disintegrate, $NP_{T}$ should reflect the increasing connectivity of the
substrate atoms to sites of a diferent kind.

\section{Results and discussion}

\subsection{Adsorption on clean surfaces}

Adsorption isotherms were simulated for the systems $Ag/Au(100)$, $%
Ag/Pt(100) $, $Au/Ag(100)$ and $Pt/Ag(100)$. The corresponding curves are
plotted in the upper part of Figs. $1$ and $2$ ($a)$ and $b)$), for
different temperatures. It can be seen that at low temperatures the
isotherms show an abrupt jump, typical of first order phase transitions. As
the temperature increases, the isotherms become smoother, specially for the
first two systems.

The present model is analogous to the Ising model in the sense that lateral
interaction between adsorbing particles is considered as a pair-potential
and only between nearest neighbors. It is well known that in this case the
critical temperature for $\theta =0.5$ can be estimated as \cite{Hill}:

\begin{equation}
T_C = \frac{J}{2 k_B ln (\sqrt{2} - 1)}
\end{equation}

\noindent where $J$ is the lateral interaction between adsorbing particles.

Taking into account the values of $J_{22}$ employed here (see table $1$),
the estimated critical temperatures are: $T_{C}=1843K$ for the system $%
Ag/Au(100)$; $T_{C}=1382K$ for $Ag/Pt(100)$; $T_{C}=3028K$ for $Au/Ag(100)$
and $T_{C}=3950K$ for $Pt/Ag(100)$. This is in agreement with the
appreciation of the isotherms in Figs. $1$ and $2$, where it can be seen
that the critical temperature must be between $1000K$ and $2000K$ for the
systems $Ag/Au(100)$ and $Ag/Pt(100)$, close to $3000K$ for $Au/Ag(100)$ and
above $3000K$ for $Pt/Ag(100).$

The chemical potential at which the isotherms of figure 1 and 2 present the
step, say $\mu _{S},$ provides a measure for the affinity of the adsorbate
for the substrate. This quantity can be substracted from the corresponding
binding energy of the adsorbate, and compared with the so-called \textit{%
underpotential shift} $\Delta \phi _{\mathbf{upd}}$, which was first defined
by Kolb et al. \cite{Kolb} as the potential difference between the
desorption peak of a monolayer of a metal $M$ adsorbed on a foreign
substrate $S$ and the current peak corresponding to the dissolution of the
bulk metal $M$. The magnitude of $\Delta \phi _{\mathbf{upd}}$ is a measure
of the affinity of the adsorbate for the substrate, as compared with the
affinity of the adsorbate with itself, and can be written in terms of the
chemical potential per particle of the atom adsorbed on a foreign substrate, 
$\mu \lbrack (S)M]$,and the chemical potential of the same species in the
bulk $\mu \lbrack (M)M],$ according to\cite{Kolb78}: 
\begin{equation}
\Delta \phi _{\mathbf{upd}}=\frac{1}{ze_{0}}(\mu \lbrack (M)M]-\mu \lbrack
(S)M])  \label{delfi0}
\end{equation}

In the present model, $\mu \lbrack (M)M]$ can be replaced by the cohesive
energy of the adsorbate, $E_{2}^{coh},$ and $\mu \lbrack (S)M]$ may be in
turn substituted by the chemical potential at which the isotherm presents
the step $\mu _{S}$, since this chemical potential correspond to the
occurrence of the adsorbate phase$.$ 
\begin{equation}
\Delta \phi _{\mathbf{upd}}^{T}(1x1)=\frac{1}{ze_{0}}(E_{2}^{coh}-\mu _{S}])
\label{delfi1}
\end{equation}%
where we have introduced the superscript $T$ to denote that this is a
theoretical prediction for the $(1x1)$ adsorbate structure . On the other
hand, the experimental estimation of the underpotential shift $\Delta \phi _{%
\mathbf{upd}}^{\exp }$ can be made from:

\bigskip 
\begin{equation}
\Delta \phi _{\mathbf{upd}}^{\exp }(1x1)=\phi (1x1)-\phi _{Nernst}
\label{delfi2}
\end{equation}%
where $\phi (1x1)$ denotes the potential at which the $(1x1)$ adsorbate
phase appears, and $\phi _{Nernst}$ is the reversible deposition potential
for the electrolyte solution employed. \ The Ag/Au(100) system\ has been
considered by different groups\cite{Ag/Au_Garcia, Ag/Au_Garcia2, Ikemiya}.
Ikemiya et al reported $\phi _{Nernst}=-58$ mV and $\phi (1x1)\approx 180mV$%
\ vs $Hg/Hg_{2}SO_{4}$ for this sytem, thus yielding $\Delta \phi _{\mathbf{%
upd}}^{\exp }(1x1)\approx 238$ $mV$. In the case of the Ag/Pt(100)system,
Aberdam et al.\cite{Aberdam} reported that the peak for the deposition of
the first monolayer of Ag was located 0.48 V more positive than the Nernst
reversible potential, so that we take this value for $\Delta \phi _{\mathbf{%
upd}}^{\exp }(1x1).$ We show in Table II these experimental values along
with the present theoretical estimations and those of our previous work
using the more accurate embedded atom method potentials. It can be
appreciated that the pair potential approximation performs almost as well as
the more sophisticated many body one. There is no evidence in the literature
for underpotential deposition of Au on Ag(100) or for Pt underpotential
deposition on Ag(100). A related system, Pt deposition on Au(100) has been
measured by Waibel et al.\cite{Waibel}, with the finding that Pt deposition
takes place at overpotentials. Since the cohesive energy of Au is larger
than that of Ag, it is expected that if Pt upd is not found on Au(100), it
will be even less probable on Ag(100).

\bigskip The lower parts of Figs. $1$ and $2$ (parts $c)$ and $d)$) show the
differential adsorption heats. The qualitative form of the curves of $q_{d}$
vs. $\mu $ is similar to the adsorption isotherms ($\theta $ vs. $\mu $).
The observed values of $q_{d}$ are in agreement with the estimation of $%
q_{d}\approx -\epsilon _{2}$ at very low coverages (corresponding to the
adsorption of particles in sites without neighbors) and $q_{d}\approx
-\epsilon _{2}-4\times J_{22}$ at high coverages (corresponding to the
adsorption of particles in sites surrounded by four nearest neighbors of the
same kind).

\subsection{Adsorption in the presence of substrate islands}

Figs. $3$ and $5$ show the adsorption isotherms for two of the four studied
systems in the presence of surface defects (substrate atoms present in the
monolayer forming islands). In the figures, the isotherms for the complete
monolayer are plotted together with the isotherms of steps and kink sites
for three different temperatures [parts $a)$, $b)$ and $c)$], as well as the
differential adsorption heats for the three temperatures [part $d)$].

It can be seen that for $Ag/Pt(100)$ (Fig. $3$) and $Ag/Au(100)$ (not
shown), the kink sites are occupied first, then the step sites and finally
the complete monolayer. For the other two systems, at $T=300K$, there is no
differentiation between the different kind of sites, but at greater
temperatures a small difference arises (see parts $a)-c)$ of Fig. $5$ for
the system $Au/Ag(100)$, the system $Pt/Ag(100)$ is not shown, but the
results are very similar).

Analyzing the behavior of the differential heat of adsorption [part $d)$],
it can be seen that for the first two systems (see Fig. $3$ for $Ag/Pt(100)$%
), at low temperatures, there are four differentiated stages, before the
adsorption of the complete monolayer (last part). The first stage with $%
q_{d}\approx -(\epsilon _{2}+2J_{12})$ (filling of substrate kink sites);
the second one with $q_{d}\approx -(\epsilon _{2}+J_{12}+J_{22})$ (filling
of step sites besides to another adsorbate particle); the third one with $%
q_{d}\approx -(\epsilon _{2}+2J_{22})$ (filling of an adsorbate kink site,
that is, adsorption on terrace sites next to two adsorbate particles during
monolayer growth) and the fourth one with $q_{d}\approx -(\epsilon
_{2}+4J_{22})$ (monolayer completion). All these stages are illustrated in
Fig. $4$. These estimations are closest to the results of the simulations
for the system $Ag/Pt(100)$.

For the other two systems (see Fig. $5$ for $Au/Ag(100)$) there are at first
sight only two more or less well defined stages: before and after monolayer
completion. In the first part, where only a few adsorbate particles come
into the system, the differential heat can be written as $q_{d}\approx
-[\epsilon _{2}+4(J_{12}-J_{11})+2J_{11}]$. This value corresponds to the
replacement of a substrate particle by an adsorbate particle in the middle
of an island and the positioning of the former at the edge of the island.
The interpretation of the value of $q_{d}$ is not straightforward for
monolayer completion.

Fig. $6$ shows the final state of a portion of the simulation cell at three
different chemical potentials for the system $Ag/Au(100)$ in the presence of
defects at two different temperatures. At $T=300K$, the islands remain
almost unchanged upon adsorbate deposition and the sites are filled
following the order: $1)$ kinks; $2)$ steps; $3)$ terraces. At $T=1000K$,
the general tendency remains but the islands show a certain disintegration.
The same behavior is found for the system $Ag/Pt(100)$ (not shown).

On the other hand, in the case of the system $Au/Ag(100)$ at $T=300K$ (upper
part of Fig. $7$), the islands do not change their shape significantly but
some adsorbate atoms penetrate inside them. At $T=1000K$ (lower part of Fig. 
$7$) the islands disintegrate completely. The same general behavior is found
for the system $Pt/Ag(100)$ (not shown).

Figs. $8$ and $9$ show the adsorption isotherms and the fraction of island
disintegration (characterized by the value $NP_{T}$, defined in equation \ref%
{NPT}) in the presence of substrate islands at three different temperatures
for two of the considered systems.

In the case of the system $Ag/Au(100)$ (Fig. $8$), it can be verified that,
at low temperatures, the islands remain unchanged, since the value of $%
NP_{T} $ is constant along the whole isotherm. As expected, the value of $%
NP_{1-0}$ decreases and the value of $NP_{1-2}$ increases as the adsorbate
atoms cover the edges of the islands. At higher temperatures, the value of $%
NP_{T}$ increases when the adsorbate atoms enter. That means that both atom
classes mix with each other, that is, the substrate islands disintegrate. At
an intermediate temperature ($T=1000K$) the value of $NP_{T}$ presents a
maximum close to the critical chemical potential due to the greater mobility
of the atoms in that situation.

For the system $Ag/Pt(100)$ (not shown) the general trend is very similar to
the previous case.

The adsorption isotherms for the systems $Au/Ag(100)$ (not shown) and $%
Pt/Ag(100)$ at different temperatures (Fig. $9a$) show that, as the
temperature increases, the jump in coverage moves towards more negative
chemical potentials. Considering the fraction of island disintegration, an
important difference with respect to the other systems consists in that even
at low temperatures the islands start to disintegrate when the adatoms are
deposited (Fig. $9b$). This is evident in the increase of the quantity $%
NP_{T}$. Another important detail to be emphasized is that for high
temperatures island disintegration is considerable, even for low adsorbate
coverages (Fig. $9d$). As anticipated in section $2.4.3$, it can be observed
in Figs. $8d$ and $9d$ that for all systems $NP_{T}\longrightarrow 0.9$ in
the limit of high temperatures and adsorbate coverage degrees.

The general picture that we get from the present results is that as long as
the susbtrate has a binding energy that is considerably larger than that of
the adsorbate, the islands of the former remain relatively unaltered upon
adsorbate formation. As the binding energy of the substrate approaches that
of the adsorbate or becomes smaller, the islands become unstable and
disintegrate relatively easily. The study of Pd ($E_{coh}=-3.91$ eV)
deposition on Au (100)($E_{coh}=-3.93$ eV)shows in fact the formation of a
surface alloy in the underpotential region\cite{Kibler}. The study of Pt ($%
E_{coh}=-5.77$ eV) deposition on Au(100) shows some remarkable features.
Although surface alloying was not reported for this system\cite{Waibel}, the
Pt deposit exhibit irregular shapes and recent molecular dynamic simulations
point towards the existence of a surface alloy\cite{Chelo}.

\section{Conclusions}

\bigskip In the present paper we revisit sytems previously simulated in
Refs. \cite{MC1, MC2} with a different model for the metal-metal
interactions. Instead of the computationally demanding many-body potentials
employed there, we use here attractive pairwise aditive potential
interactions between nearest neighbors with only a minimal set of
parameters. While the main results remain qualitatively similar, the present
modelling is considerably simple and workable in the framework of the
pairwise additive potential lattice model. It also saves considerable
computational time. Furthermore, we have considered here the influence of
the temperature and some new quantities, like differential heat and fraction
of island disintegration, that helped us in the understanding of the
simulations.

The present formulation shows that the main characteristics to be taken into
account is the strenght of the interactions between adsorbate atoms as
compared with the interaction between substrate atoms. In this sense, we can
divide the four systems studied as examples into two groups. The first group
is integrated by the systems $Ag/Au(100)$ and $Ag/Pt(100)$. In these cases
the interactions between adsorbate atoms is weaker than the interaction
between substrate atoms (i.e.,$|J_{22}|<|J_{11}|$). The second group is
integrated by the systems $Au/Ag(100)$ and $Pt/Ag(100)$, where the
interactions between adsorbate atoms is stronger than the interaction
between substrate atoms (i.e., $|J_{22}|>|J_{11}|$).

For defect-free surfaces, the adsorption isotherms show an abrupt jump at
low temperatures and become smoother at high temperatures. This is
indicative of the existence of a first order phase transition and a critical
temperature related with the interaction between adsorbate atoms.

The adsorption in the presence of islands of the same nature as that of the
substrate on the surface was also studied.

For systems in which the interaction between adsorbate particles is weaker
than the interaction between substrate particles ($Ag/Au(100)$ and $%
Ag/Pt(100)$), the substrate islands remain relatively unchanged at low
temperatures and show a certain degree of mobility at high temperatures. For
systems in which the interaction between adsorbate particles is stronger
than the interaction between substrate particles ($Au/Ag(100)$ and $%
Pt/Ag(100)$), the adsorbate atoms penetrate into the islands at low
temperatures and the islands are completely disintegrated at high
temperatures.

The simplicity of the present formulation will allow the analysis of problem
at hand in terms of a few parameters than can be systematically varied. In
other words, instead of system-oriented simulations, studies with these
control parameters can be performed, with the consequent gain of generality.

\section{Acknowledgements}

Financial support from CONICET, Agencia C\'{o}rdoba Ciencia, Secyt U.N.C.,
Program BID 1201/OC-AR PICT N%
%TCIMACRO{\U{ba} }%
%BeginExpansion
${{}^o}$
%EndExpansion
%
%
%
%
%
%
%
%
%
%
%
% %BeginExpansion ${{}^o}$ %EndExpansion
06-12485 is gratefully acknowledged. M.C. Gim\'{e}nez thanks CONICET for a
postdoctoral fellowship.

%%%%                          Bibliografia

\newpage

\section{Figure Captions}

\textbf{Figure 01} $a)$ Adsorption isotherms at different temperatures for
the system $Ag/Au(100)$. $b)$ Adsorption isotherms at different temperatures
for the system $Ag/Pt(100)$. $c)$ Differential heats for the adsorption
isotherms of the system $Ag/Au(100)$. $d)$ Differential heats for the
adsorption isotherms of the system $Ag/Pt(100)$.

\textbf{Figure 02} $a)$ Adsorption isotherms at different temperatures for
the system $Au/Ag(100)$. $b)$ Adsorption isotherms at different temperatures
for the system $Pt/Ag(100)$. $c)$ Differential heats for the adsorption
isotherms of the system $Au/Ag(100)$. $d)$ Differential heats for the
adsorption isotherms of the system $Pt/Ag(100)$.

\textbf{Figure 03} $a)-c)$ Adsorption isotherms in the presence of surface
defects for the complete monolayer, the step sites and the kink sites at
three different temperatures for the system $Ag/Pt(100)$. $d)$: differential
heat of adsorption at the three temperatures for the same system.

\textbf{Figure 04} Schematic representation of the top view of four
environment types close to the adatom adsorption site. $a)$: substrate kink
site. $b)$: substrate step site, next to another adsorbate particle. $c)$:
adsorbate kink site. $d) $: hollow site (upon monolayer completion). Filled
circles denote substrate atoms. Empty circles represent adsorbate atoms. The
empty dashed circle denotes the adsorption site under consideration. The
underlying substrate atoms are not shown.

\textbf{Figure 05} $a)-c)$Adsorption isotherms in the presence of surface
defects for the complete monolayer, the step sites and the kink sites at
three different temperatures for the system $Au/Ag(100)$. $d)$: differential
heat of adsorption at the three temperatures for the same system.

\textbf{Figure 06} Snapshots showing the final state of the surface at
different chemical potentials for the system $Ag/Au(100)$ in the presence of
surface defects at $T = 300 K$ and $T = 1000 K$. Filled circles represent
gold atoms while non-filled ones represent silver atoms.

\textbf{Figure 07} Snapshots showing the final state of the surface at
different chemical potentials for the system $Au/Ag(100)$ in the presence of
surface defects at $T = 300 K$ and $T = 1000 K$. Filled circles represent
silver atoms while non-filled ones represent gold atoms.

\textbf{Figure 08} . $a)$Adsorption isotherms in the presence of surface
defects for the complete monolayer at three different temperatures for the
system $Ag/Au(100)$. $b)-d)$ Fraction number of pairs ($NP_{1-0}$, $NP_{1-2}$
and $NP_{T}$) at the same temperatures.

\bigskip \textbf{Figure 09} $a)$: Adsorption isotherms in the presence of
surface defects for the complete monolayer at three different temperatures
for the system $Pt/Ag(100)$. $b)-d)$ Fraction number of pairs ($NP_{1-0}$, $%
NP_{1-2}$ and $NP_{T}$) at the same temperatures.

\newpage

%%%%%%%%%%%%%%%%%            TABLAS %%%%%%%%%%%%%%%%%%%%%%%%%%%%%%%

\section{Tables}

\begin{center}
\begin{tabular}{|c|c|c|c|c|c|}
\hline
$System$ & $\epsilon_1$ & $\epsilon_2$ & $J_{11}$ & $J_{22}$ & $J_{12}$ \\ 
\hline\hline
$Ag/Au(100)$ & $-3.05$ & $-2.58$ & $-0.54$ & $-0.28$ & $-0.42$ \\ \hline
$Ag/Pt(100)$ & $-4.34$ & $-3.13$ & $-0.83$ & $-0.21$ & $-0.56$ \\ \hline
$Au/Ag(100)$ & $-2.39$ & $-3.11$ & $-0.25$ & $-0.46$ & $-0.36$ \\ \hline
$Pt/Ag(100)$ & $-2.30$ & $-4.22$ & $-0.17$ & $-0.60$ & $-0.41$ \\ \hline
\end{tabular}
\end{center}

\textbf{Table 1}: Parameters representing the adsorption energies and the
interaction energies (in $eV$ units) employed here for the considered
systems.

\bigskip

\bigskip 
\begin{tabular}{llll}
System & $\Delta \phi _{\mathbf{upd}}^{T}(1x1)$ $/$ $V$ & $\Delta \phi _{%
\mathbf{upd}}^{T}(1x1),EAM$ $/$ $V$ & $\Delta \phi _{\mathbf{upd}}^{\exp
}(1x1)$ $/$ $V$ \\ 
$Ag/Au(100)$ & 0.23 & 0.17 & 0.24$^{a}$ \\ 
$Ag/Pt(100)$ & 0.67 & 0.55 & 0.48$^{b}$ \\ 
$Au/Ag(100)$ & 0.04 & -0.08 & < 0?$^{c}$ \\ 
$Pt/Ag(100)$ & -0.44 & -0.53 & < 0$?^{c}$%
\end{tabular}

\textbf{Table 2: }Calculated and experimental underpotential shifts for the
systems considered in the present work. $\Delta \phi _{\mathbf{upd}%
}^{T}(1x1) $ denote results of this work, $\Delta \phi _{\mathbf{upd}%
}^{T}(1x1)$ $/$ $EAM$ are results from previous simulations using the
many-body potential from the embedded atom method\cite{Daw-Baskes}, $\Delta
\phi _{\mathbf{upd}}^{\exp }(1x1)$ are experimental estimations taken from
the literature.$\Delta \phi _{\mathbf{upd}}^{T}(1x1)$ values were calculated
using equation (\ref{delfi1}), with $E_{2}^{coh}$= --2.85, -3.93 and -5.77
eV for Ag, Au and Pt respectively.

a)taken from reference \cite{Ikemiya}, b)Taken from reference \cite{Aberdam}
c)no upd has been reported in the literature for these systems so far.

\bigskip

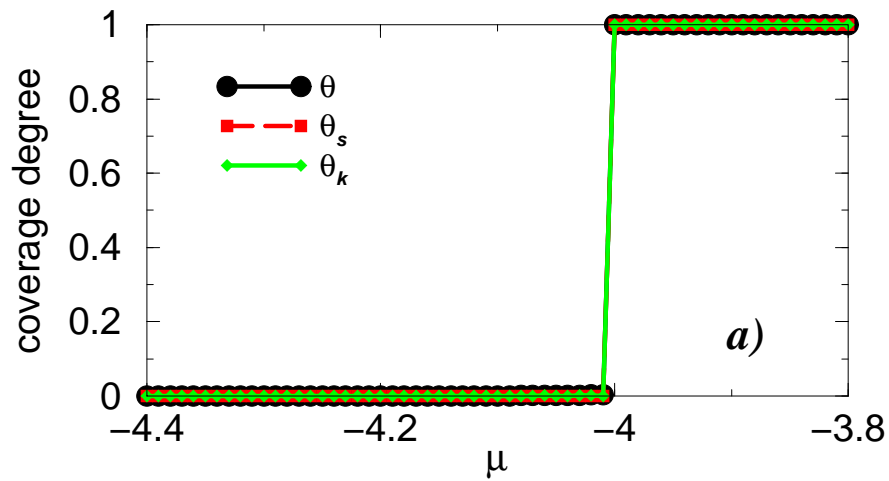
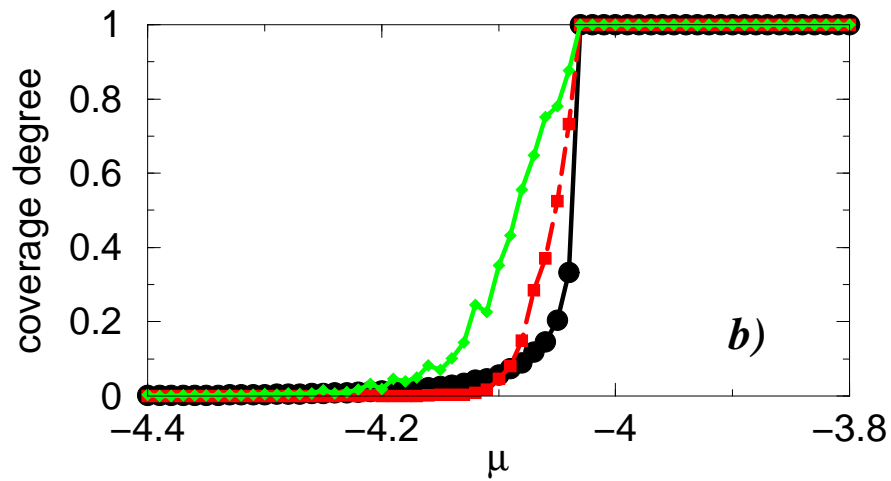
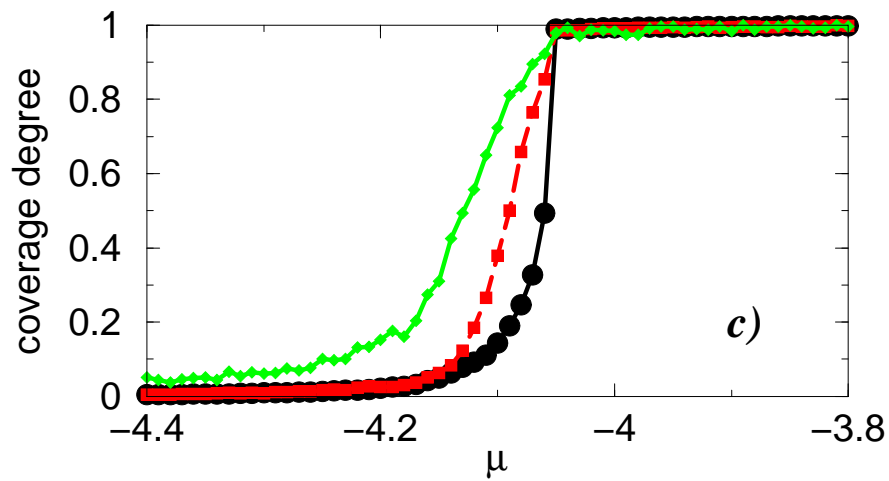
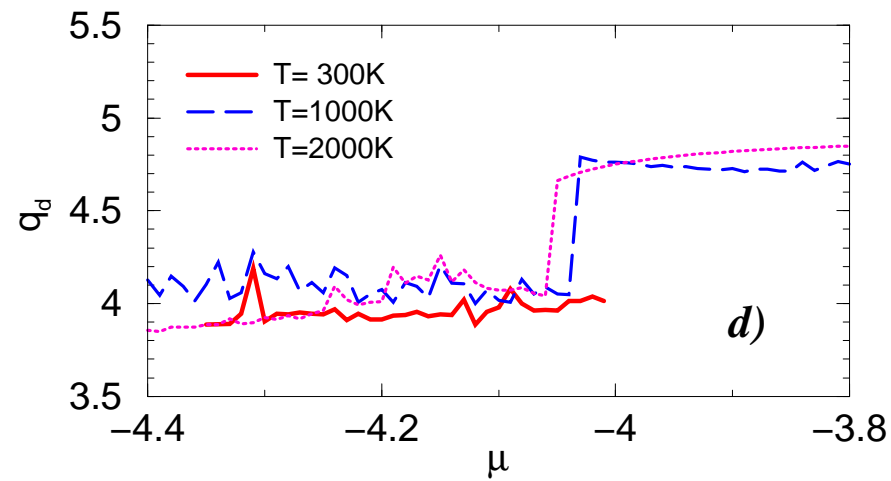
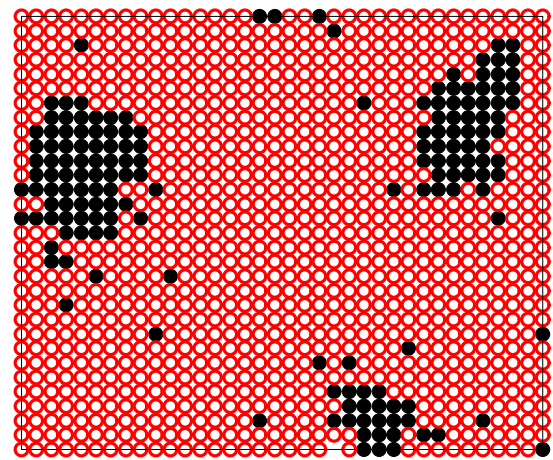
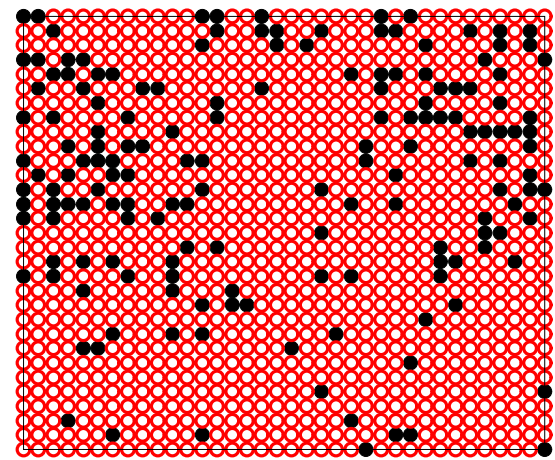
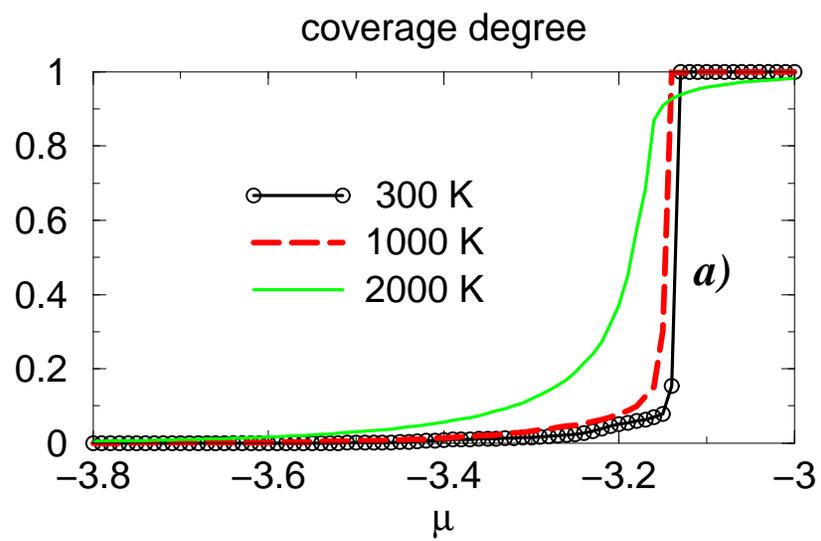
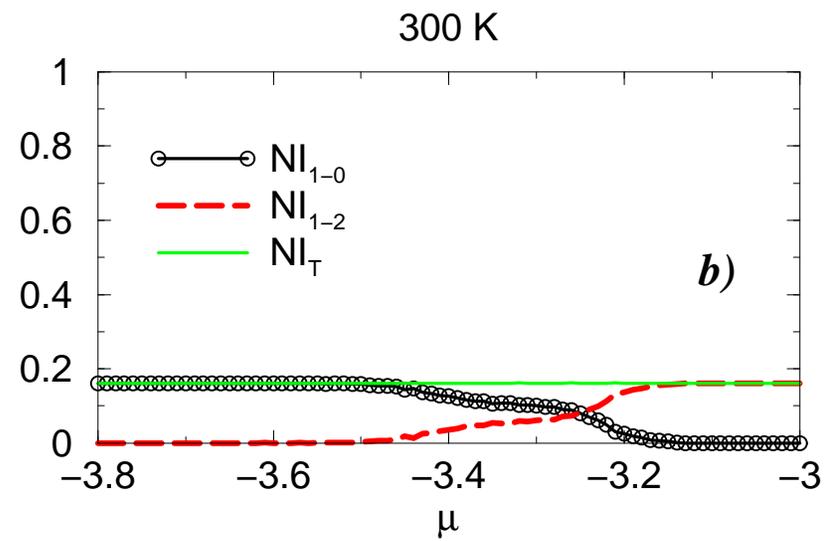
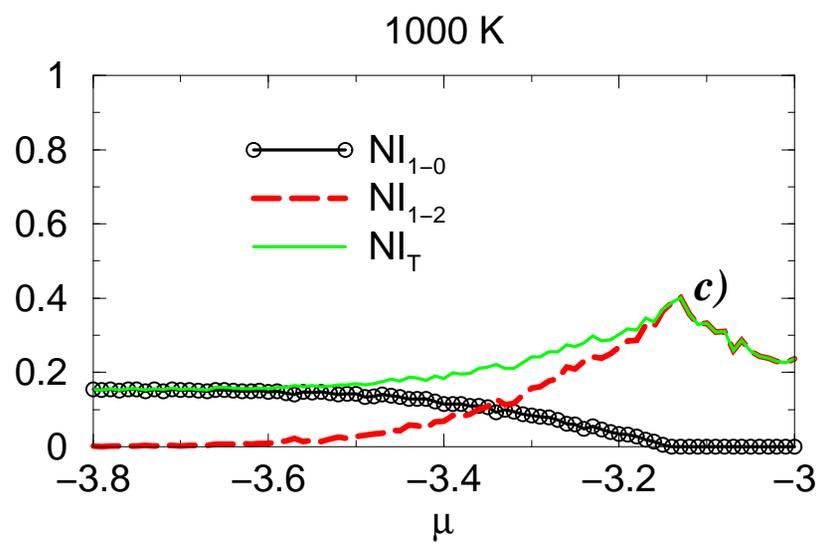
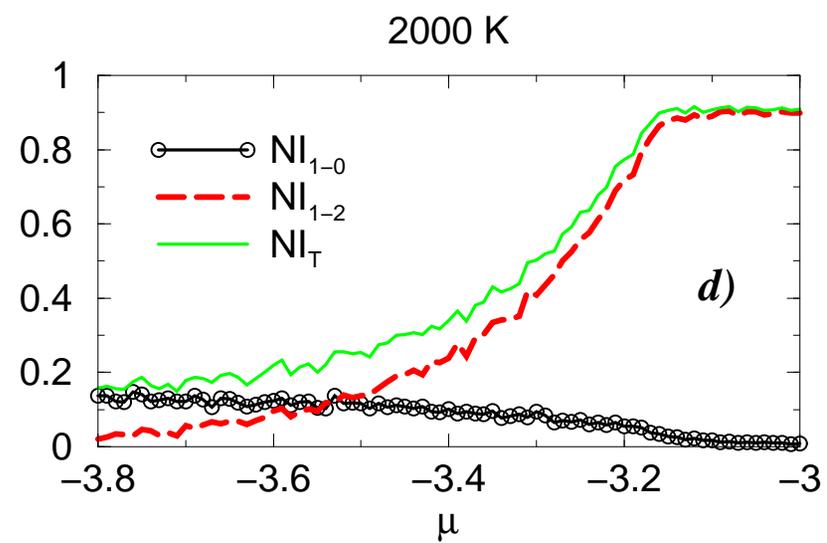
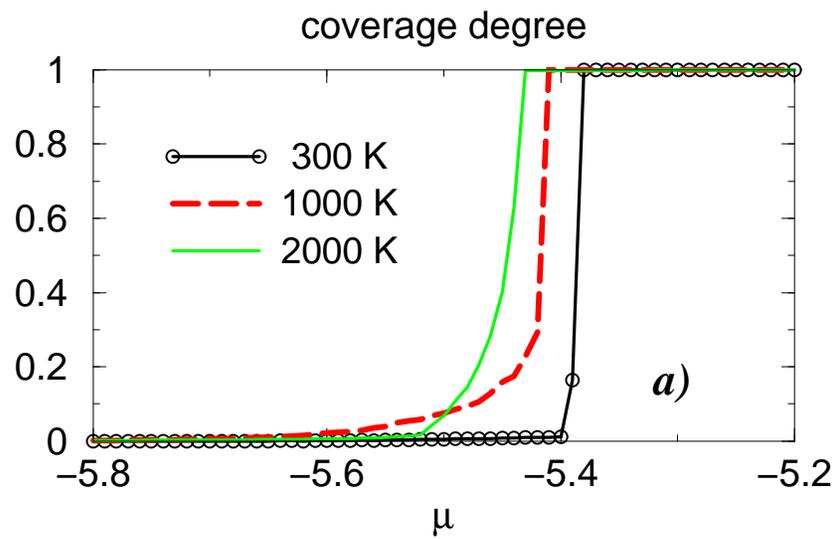
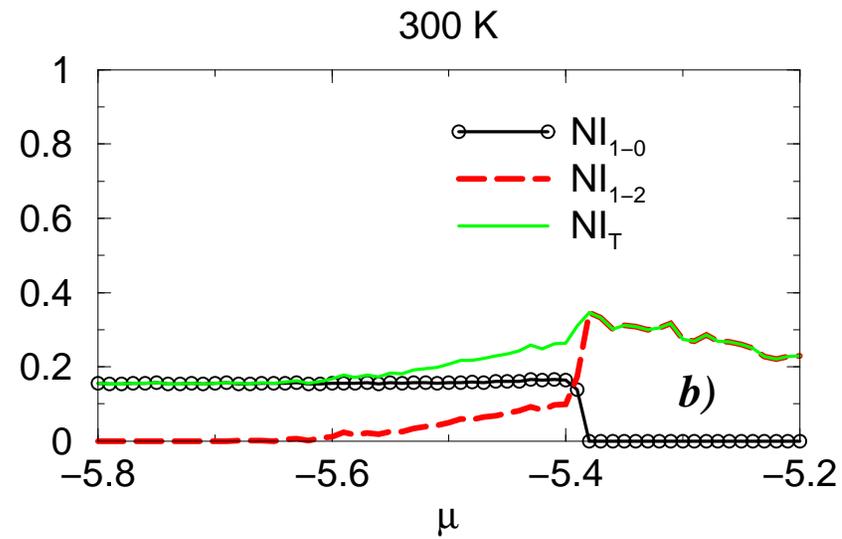
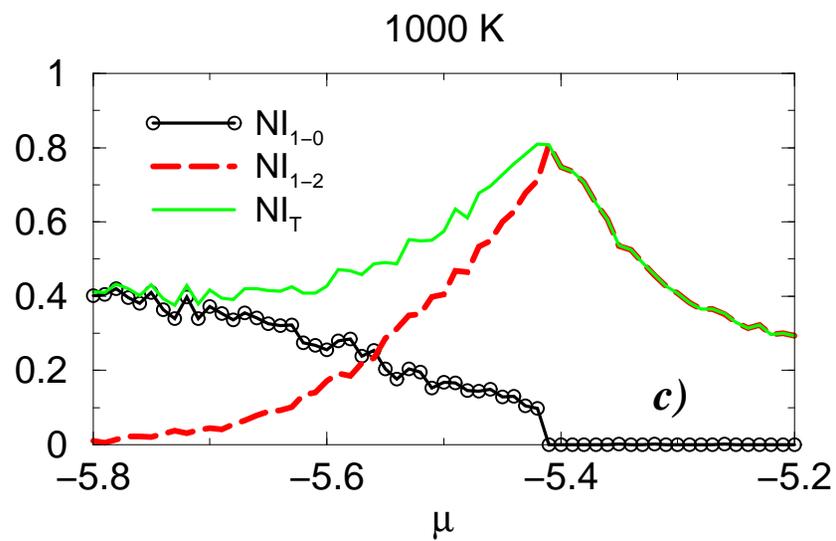
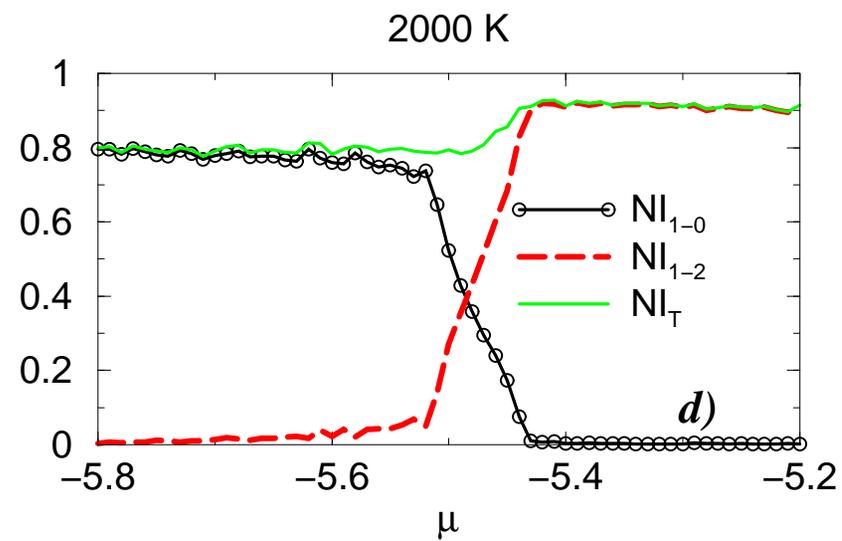
\end{document}